\documentclass[onecollarge,natbib]{svjour2}
\bibpunct{[}{]}{;}{n}{}{,} 
\smartqed  
\usepackage{graphicx}
\usepackage[shortcuts]{extdash} 
\usepackage{amsmath}
\usepackage{amssymb}

\usepackage{MnSymbol} 
\journalname{Few-Body Systems}
\begin{document}

\title{Single particle momentum distributions for three-bosons in two and three 
dimensions and dimensional crossover}

\author{F. F. Bellotti         \and
        M. T. Yamashita
}

\institute{F.~F. Bellotti \at
              Department of Physics and Astronomy, Aarhus University, DK-8000 Aarhus C, Denmark \\
              Instituto Tecnol\'{o}gico de Aeron\'autica, 12228-900, S\~ao Jos\'e dos Campos, SP, Brazil \\
              Instituto de Fomento e Coordena\c{c}{\~a}o Industrial, 12228-901, S{\~a}o Jos{\'e} dos Campos, SP, Brazil \\
              Tel.: +45 87 15 55 95\\
              Fax: +45 86 12 07 40\\
              \email{bellotti@phys.au.dk}             \\
           \and
           M. T. Yamashita \at
             Instituto de F\'\i sica Te\'orica, UNESP-Univ Estadual Paulista, C.P. 70532-2, 
             S\~ao Paulo, CEP 01156-970, Brazil
}

\date{Received: date / Accepted: date}

\maketitle

\begin{abstract}
In this paper, we review our main results involving the single particle momentum 
distribution of bosonic trimer states in two and three dimensions. A summary table 
makes easier the comparison between the matrix elements and the different terms of the 
momentum distributions. We also show a practical method to continuously interpolate 
between different dimensions.
\end{abstract}

\section{Introduction}
\label{intro}

Many nature laws can be strongly affected if dimensionality is changed. As already pointed out by 
Landau in his classic book \cite{landau1977}, any infinitesimal amount 
of attraction produce a bound state in 2D, while a finite amount of attraction is necessary 
to bind a 3D system. A remarkable phenomenon related to the dimensionality of the system 
arises in the study of three identical bosons, where the differences in the energy spectrum (and 
also other observables) are directly related to the number of dimensions 
that this system may access: in 2D there are only two three-body bound states linked to 
one two-body bound state in the limit where the range of the potential goes to 
zero \cite{bruchPRA1979}. On the other hand, in 3D, the number of three-body bound states may grow 
to infinity \cite{efimovYF1970,thomasPR1935} - this effect is now called by Efimov effect.

The Efimov effect corresponds to an accumulation of the three-boson energy levels, toward zero 
energy, when the two-body scattering length tends to infinity. In this limit, where the two-body 
energy is zero, the energies of successive states are geometrically spaced obeying a universal 
ratio. These states were predicted and observed for three identical bosons in 3D 
systems \cite{kraemerNP2006,ferlainoPOJ2010}, but are absent in 2D even in the most favorable 
scenario of mass\=/imbalanced systems \cite{limZfPAHaN1980,adhikariPRA1988}, where a 
mass\=/dependent effective potential favors the binding of a light particle to a heavy 
dimer \cite{bellottiJoPB2013}.

The appearance of Efimov states in 3D is very closely related to the possibility of collapse 
the three-body system. This collapse (Thomas collapse), firstly derived by Thomas in 1935, says 
that the three-body ground state energy may be made as deep as you want by decreasing the 
range of the potential ($r_0$) - in the limit $r_0\rightarrow 0$ the three-body binding 
energy tends to infinity. This divergence demands the inclusion of a cutoff or, equivalently, 
a new physical scale independent of the two-body energy. In 2D, this collapse is absent in 
such a way the three-body observables are proportional to the two-body energy. For example, the three-body ground 
state energy is 16.52$E_2$ and the energy of the first excited state is 1.267$E_2$ for three identical bosons\cite{bruchPRA1979}.

There are many examples of observables in cold atomic gases that are affected by the 
dimensionality of the system. We would like to start mentioning the two- ($C_2$) and 
three-body ($C_3$) contact parameters. The connection between universal two\=/body 
correlations to many\=/body properties through the quantity $C_2$ was proposed by Tan 
in \cite{tanAoP2008} (this quantity is often called Tan's contact parameter). 
For example, the variation in the energy of a Fermi gas of momentum $k_F$ with the 
interaction strength (scattering length $a$) is directly proportional to this $C_2$, 
namely
\begin{equation}
2 \pi \frac{d E}{d\left[-1/(k_F a) \right]}= C_2 \; .
\label{energyc2}
\end{equation}
Furthermore, the virial theorem for this atomic gas also relates with $C_2$ through
\begin{equation}
E-2V=-\frac{C_2}{4 \pi k_F a} \; .
\label{virialc2}
\end{equation} 
These relations, in the way they are presented, were confirmed in experiments with two-component 
Fermi gases \cite{kuhnlePRL2010}, where each side of Eqs.~\eqref{energyc2} and \eqref{virialc2} 
were measured independently and after compared to each other. A later experiment showed that 
similar relations also hold for bosons \cite{wildPRL2012}.

The quantities on the left-hand-side of Eqs.~\eqref{energyc2} and \eqref{virialc2} are defined through the many\=/body properties of the gas, while the contact parameter is defined in the few\=/body sector. A way to determine this parameter is to find the coefficient in the leading order of the asymptotic one-body large momentum density, $n(q)$, of few-body systems, given by
\begin{equation}
\lim_{q \to \infty} n(q) \rightarrow \frac{C_2}{q^4}+C_3 F(q)+... \; .
\label{asymomem}
\end{equation} 
The next order in this expansion defines the three-body contact parameter, $C_3$, which may be important only for bosonic systems, since the Pauli principle suppresses the short-range correlations for two-component Fermi gases. Notice that the momentum dependence of the leading order term in this expansion is the same for 1D, 2D and 3D systems \cite{valientePRA2012}, but the function $F(q)$ depends on the dimensionality of the system \cite{bellottiPRA2013}. The two- and three-body contact parameters were determined for three identical bosons in 2D \cite{bellottiPRA2013} and 3D \cite{castinPRA2011}, and for mixed-species systems in 2D \cite{bellottiNJoP2014} and 3D \cite{yamashitaPRA2013}.

An interesting point about the two-body contact parameter is that, despite the considerable difference between the binding energy of both states for three identical bosons in 2D (where the well-known limit cycle is not present), the ratio $\frac{C_2}{E_3}$ is the same for the two states \cite{bellottiPRA2013}. In general, for a mixed-species system - which have a richer energy 
spectrum \cite{bellottiJoPB2013,bellottiPRA2012} - the ratio $\frac{C_2}{E_3}$ is not the 
same for all states, but only in the special and experimentally accessible case of a 
three-body system composed for at least two identical non-interacting particles \cite{bellottiNJoP2014}.

Among the several differences involving the dimensionality of the system we would like 
to stress that the function $F(q)$ in Eq.~\eqref{asymomem} has very distinct forms in each 
case of 2D or 3D. This function is directly related to the spectator functions $f(q)$ (given in Eqs.~\eqref{wave} and \eqref{spec}), whose asymptotic 
form were discovered in the 60's for 3D systems \cite{danilovZETF1961} and approximately 50 years 
later for 2D systems \cite{bellottiPRA2013,bellottiNJoP2014}. Thus, an interesting question is whether it is possible to interpolate between the 3D and 2D limits in a simple theoretical way 
and subsequently explore this in simulations using both more involved numerical methods and 
experimental setups since the development of the techniques for cooling and trap atoms allows 
the interpolation between different dimensions~\cite{blochRMP2008,lamporesiPRL2010}.

We proposed a model that has the ability to interpolate geometrically between two and three spatial 
dimensions and thus study this crossover for both two- and three-body bound states 
of identical bosons. A ``squeezed'' dimension, whose size can be varied to interpolate the two 
limits, is employed with periodic boundary conditions (PBC). This model has the unique feature 
that it can be regularized analytically, which is a great advantage for its numerical implementation allowing to go smoothly between both limits. The theoretical elegance and 
tractability of calculations in the three-body system is itself a strong incentive for pursuing 
this geometry, but in spite of this elegance, a direct connection between experiments and the 
parameter that dials between different dimensions with PBC in this model was not found yet.

In the next sections we review our main results involving the momentum distributions in three and 
two dimensions. Both results are put together side-by-side in a table where the comparison becomes 
easier. In the last section we give an overview of our method that continuously interpolates 
between 3 and 2D limits.

\section{Integral equation for bound states}
\label{sec:1.1}
We investigate $abc$ bound systems whose dynamics is restricted to either two (2D) or three spatial dimensions (3D). The masses are $m_a,m_b,m_c$ and the pairwise interactions are described for attractive zero-range potentials, being $E_{ab},E_{ac},E_{bc}$ the energy of each pair. The three-body wave function $\langle \mathbf{q}_\alpha,\mathbf{p}_\alpha|\Psi_{abc}\rangle$ has the same functional form in both 2D and 3D. For any $s-$wave bound state, the energy $E_3$ is a solution of the free Schr\"odinger equation, except in the region where particles overlap. Using the Faddeev decomposition in momentum space, the bound state wave function in units of $\hbar=1$ is written as 
\begin{equation}
\left\langle \mathbf{q}_\alpha,\mathbf{p}_\alpha \right.\left|\Psi_{abc}\right\rangle=\Psi\left(\mathbf{q}_\alpha,\mathbf{p}_\alpha\right)=\frac{f_{\alpha}\left(q_\alpha\right)+f_{\beta}\left(\left| \mathbf{p}_\alpha- \frac{m_\beta}{m_\beta+m_\gamma}\mathbf{q}_\alpha\right|  \right)+f_{\gamma}\left(\left| \mathbf{p}_\alpha+ \frac{m_\gamma}{m_\beta+m_\gamma}\mathbf{q}_\alpha\right| \right)}{|E_{3}|+\frac{q_\alpha^{2}}{2m_{\beta \gamma,\alpha}}+\frac{p_\alpha^{2}}{2m_{\beta \gamma}}} , 
\label{wave}
\end{equation}
where $\alpha,\beta,\gamma$ are cyclic permutations of $a,b,c$, $\mathbf{q}_\alpha$ is the $\alpha$ particle momenta with respect to the CM of the pair $\beta\gamma$, $\mathbf{p}_\alpha$ is the pair relative momenta, $m_{\beta \gamma,\alpha}= m_\alpha(m_\beta+m_\gamma)/(m_\alpha+m_\beta+m_\gamma)$ and $m_{\beta \gamma}= (m_\beta+m_\gamma)/(m_\beta+m_\gamma)$ are the reduced masses and $f_{\alpha,\beta,\gamma}(\mathbf{q})$ are the Faddeev components, or spectator functions.
The three-body energy, $E_3$, and the spectator functions $f_{\alpha,\beta,\gamma}$ are solution of a set of three coupled homogeneous integral equations, which in a compact form reads
\begin{equation}
f_{\alpha}\left( \mathbf{q}\right)  =\tau^{D}_\alpha(q,E_3) \int_0^\infty k\;dk \left[  K^{D}_{\alpha \beta}(q,k,E_3) \; f_{\beta}\left(k\right)  + K^{D}_{\alpha \gamma}(q,k,E_3) \; f_{\gamma}\left(k\right) \right] ,
\label{spec}
\end{equation}
where the matrix elements of the two-body T-matrix, $\tau^{D}_\alpha(q,E_3)$ and the kernels $K^{D}_{\alpha \beta}(q,k,E_3)$ and $K^{D}_{\alpha \gamma}(q,k,E_3)$ are given in table~\ref{tab:1}. 

An interesting difference between 2D and 3D three-body systems can be seen in Eq.~\eqref{spec} and table~\ref{tab:1}.
For each non-interacting $\beta \gamma$ pair, the respective spectator function $f_\alpha(q)=0$. Choosing $E_{\beta \gamma}=0$ in the first line of table~\ref{tab:1} gives different result for 2D and 3D systems. In 3D,  $\left[\tau^{D}_\alpha(q,E_3)\right]^{-1}$ is finite and Eq.~\eqref{spec} is well-defined even if the three two-body subsystems interact with zero energy. On the other hand in 2D, $\left[\tau^{D}_\alpha(q,E_3)\right]^{-1} \to \infty$, meaning that if at least two pairs have zero energy, Eq.~\eqref{spec} is not well\=/defined and three\=/body bound states do not exist. Therefore, non-interacting and zero-energy two-body systems lead to the same result in 2D, while they can give completely different results in 3D~\cite{yamashitaPRA2013}.
\begin{table}[htb!]
\caption{Matrix elements of the two-body T-matrix, $\tau^{D}_\alpha(q,E_3)$ and the kernels $K^{D}_{\alpha \beta}(q,k,E_3)$ and $K^{D}_{\alpha \gamma}(q,k,E_3)$ for both 2D and 3D systems, where $\mu$ is the subtraction point (see, for example, \cite{fredericoPiPaNP2012}).} 
\centering
\label{tab:1}       
\begin{tabular}{lcc}
\hline\noalign{\smallskip}
  & 2D & 3D   \\[3pt]
\tableheadseprule\noalign{\smallskip}
$\left[\tau^{D}_\alpha(q,E_3)\right]^{-1}$ & 
$4\pi m_{\beta \gamma}\ln \left( \sqrt{\frac{\frac{q^{2}}{2m_{\beta \gamma,\alpha} }-E_{3}}{|E_{\beta\gamma}|}}\right)$ & 
$\pi  \left( 2 m_{\beta \gamma} \right)^{3/2} \left( \sqrt{ \left(\frac{q^{2}}{2m_{\beta \gamma,\alpha} }-E_{3} \right)} - \sqrt{|E_{\beta \gamma}| }\right)$ \\ \noalign{\smallskip}
$K^{D}_{\alpha \beta}(q,k,E_3)$ & 
$\frac{ 1 }{\sqrt{\left(-E_{3}+\frac{q^{2}}{2 m_{\alpha \gamma}}  +\frac{k^{2}}{2 m_{\beta \gamma}}\right)^2-\left(\frac{k \; q} {m_\gamma}\right)^2}}$  & 
$ \frac{m_\gamma }{q}\left(\ln \frac{-E_{3}+\frac{q^{2}}{2 m_{\alpha \gamma}}+\frac{k^{2}}{2 m_{\beta \gamma}}+\frac{k\;q}{m_\gamma} }{-E_{3}+\frac{q^{2}}{2 m_{\alpha \gamma}}+\frac{k^{2}}{2 m_{\beta \gamma}}-\frac{k\;q}{m_\gamma} } - \ln \frac{\mu^2+\frac{q^{2}}{2 m_{\alpha \gamma}}+\frac{k^{2}}{2 m_{\beta \gamma}}+\frac{k\;q}{m_\gamma} }{\mu^2+\frac{q^{2}}{2 m_{\alpha \gamma}}+\frac{k^{2}}{2 m_{\beta \gamma}}-\frac{k\;q}{m_\gamma} } \right)$  \\ \noalign{\smallskip}
$K^{D}_{\alpha \gamma}(q,k,E_3)$ &
$\frac{1 }{\sqrt{\left(-E_{3}+\frac{q^{2}}{2m_{\alpha \beta }}+\frac{k^{2}}{2m_{\beta \gamma}}\right)^2-\left(\frac{k \; q}{m_\beta}\right)^2}}$&
$ \frac{m_\beta }{q} \left(\ln \frac{-E_{3}+\frac{q^{2}}{2m_{\alpha \beta }}+\frac{k^{2}}{2m_{\beta \gamma}}+\frac{k\;q}{m_\beta}}{-E_{3}+\frac{q^{2}}{2m_{\alpha \beta }}+\frac{k^{2}}{2m_{\beta \gamma}}-\frac{k\;q}{m_\beta}} - \ln \frac{\mu^2+\frac{q^{2}}{2m_{\alpha \beta }}+\frac{k^{2}}{2m_{\beta \gamma}}+\frac{k\;q}{m_\beta} }{\mu^2+\frac{q^{2}}{2m_{\alpha \beta }}+\frac{k^{2}}{2m_{\beta \gamma}}-\frac{k\;q}{m_\beta}} \right)$\\
\noalign{\smallskip}\hline
\end{tabular}
\end{table}

\section{Momentum distribution}
\label{sec:1.2}

The one-body density functions are observable quantities even in the limit of large momenta where the number of atoms is small, which has already been observed in experiments using time-of-flight and the mapping to momentum space \cite{stewartPRL2010}, 
Bragg spectroscopy \cite{kuhnlePRL2010} or momentum-resolved photo-emission spectroscopy \cite{frohlichPRL2011}.

The one-body momentum density of the particle $\alpha$ is defined through the wave function $\Psi(\mathbf{q}_\alpha,\mathbf{p}_\alpha)$ from Eq.~\eqref{wave} as
\begin{equation}
n(q_\alpha)=\int{d^{D} p_\alpha  |\Psi(\mathbf{q}_\alpha,\mathbf{p}_\alpha)|^2}
\label{dist}
\end{equation}
and  the normalization is $\int{d^{D}  q_\alpha\;n(q_\alpha)}=1$, where $D=2,3$ for 2D or 3D systems, respectively.  
  
Inserting Eq.~\eqref{wave} in Eq.~\eqref{dist} and expanding it, the nine initial terms can be grouped into four components by using arguments of symmetry, each one with a distinctly 
different integrand structure.  The
one-body momentum density is expressed as a sum of this four terms, i.e.,
$n(q_\alpha)=\sum_{i=1}^4{n_i(q_\alpha)}$.

A general system of three distinguishable particles, presents three
distinct one-body momentum distributions, each one corresponding to a
different particle.  The four terms for particle $\alpha$ are expressed as
\begin{align}
n_1(q_\alpha)&=\left|f_{\alpha}\left(q_\alpha\right)\right|^2  \int{d^{D} k \frac{1}{\left(-E_{3}+\frac{q_\alpha^{2}}{2m_{\beta \gamma,\alpha}}+\frac{k^{2}}{2m_{\beta \gamma}}\right)^2}}  , 
\label{eqch5.03a}\\
n_2(q_\alpha)&= \int{d^{D} k\frac{\left|f_{\beta}(k)\right|^2}{\left(-E_{3}+\frac{q_\alpha^{2}}{2m_{\alpha \gamma}}+\frac{k^{2}}{2m_{\beta \gamma}}+\frac{\mathbf{k}\cdot \mathbf{q_\alpha}}{m_\gamma}\right)^2}}  +
\int{d^{D} k\frac{\left|f_{\gamma}(k)\right|^2}{\left(-E_{3}+\frac{q_\alpha^{2}}{2m_{\alpha \beta}}+\frac{k^{2}}{2m_{\beta \gamma}}-\frac{\mathbf{k}\cdot \mathbf{q_\alpha}}{m_\beta}\right)^2}} \; , 
\label{eqch5.03b}\\
n_3(q_\alpha)&=  f^\ast_{\alpha}\left(q_\alpha\right) \int d^{D} k \left[\frac{  f_{\beta}(k)}{\left(-E_{3}+\frac{q_\alpha^{2}}{2m_{\alpha \gamma}}+\frac{k^{2}}{2m_{\beta \gamma}}+\frac{\mathbf{k}\cdot \mathbf{q_\alpha}}{m_\gamma}\right)^2} 
+  \frac{ f_{\gamma}(k)}{\left(-E_{3}+\frac{q_\alpha^{2}}{2m_{\alpha \beta}}+\frac{k^{2}}{2m_{\beta \gamma}}-\frac{\mathbf{k}\cdot \mathbf{q_\alpha}}{m_\beta}\right)^2} \right] + c.c. \; ,  
\label{eqch5.03c}\\
n_4(q_\alpha)&=  \int{d^{D} k\frac{f_{\beta}\left( \left|\mathbf{k}-\frac{m_\beta}{m_\beta+m_\gamma}\mathbf{q_\alpha}\right|\right) f^\ast_{\gamma}  \left( \left|\mathbf{k}+\frac{m_\gamma}{m_\beta+m_\gamma}\mathbf{q_\alpha}\right|\right)}{\left(-E_{3}+\frac{q_\alpha^{2}}{2m_{\beta \gamma,\alpha}}+\frac{p^{2}}{2m_{\beta \gamma}}\right)^2}} + c.c. \; . 
\label{eqch5.03d}
\end{align}
Notice that the distributions for the other particles are obtained by cyclic
permutations of $(\alpha,\beta,\gamma)$ in these expressions. 

Although the equations for bound states (Eq.~\eqref{spec}) and momentum distributions (Eqs.~\eqref{eqch5.03a} to \eqref{eqch5.03d}) were derived for a general case of three distinguishable particles, we now specialize to experimentally relevant systems composed by two identical bosons $a$ and a distinct particle $b$. The large momentum limit of Eqs.~\eqref{eqch5.03a} to \eqref{eqch5.03d} were derived in detail for 2D and 3D systems respectively in \cite{bellottiNJoP2014,yamashitaPRA2013} and the final result for the density profile of particle $b$ with respect to the pair $aa$ is presented in table~\ref{tab:2}, in units of $m_a=1$. 
\begin{table}[htb!]
\caption{Asymptotic forms for both 2D and 3D systems where ${\cal A}=\frac{m_b}{m_a}$, $\tan\theta_3=\sqrt{\frac{{\cal A}+2}{{\cal A}}}$ for $0\leq\theta_3\leq\pi/2$ and $\tan\theta_4=\sqrt{{\cal A}({\cal A}+2)}$ for $0\leq\theta_4\leq\pi/2$, $\Gamma$, $c_a$ and $c_b$ are normalization constants and $n_5(q_b)$ is the second order term in the expansion of $n_2(q_b)$  in Eq.~\eqref{eqch5.03b}.}
\centering
\label{tab:2}  
\begin{tabular}{c c c}
\hline\noalign{\smallskip}
  & 2D & 3D   \\[3pt]
\tableheadseprule\noalign{\smallskip}
$n_1(q_b)$ & 
$16 \pi \frac{{\cal A}}{{\cal A}+2} \Gamma^2 \frac{\ln^2 (q_b)}{q_b^6}$ & 
$\frac{\pi^2 \left|c_{b}\right|^2}{q_b^5} \sqrt{ \frac{{\cal A}}{{\cal A}+2}}$ \\ \noalign{\bigskip}
$n_2(q_b)$ & 
$\frac{16 \pi}{q_b^4} \frac{{\cal A}^2}{\left( {\cal A}+1\right)^2} \int_0^\infty {dk\;k\;\left|f_{a}(k)\right|^2}$ & 
$ \frac{32 \pi}{q_b^4} \frac{{\cal A}^2}{({\cal A}+1)^2} \int_0^\infty dk\; k \;\left|f_{a}(k)\right|^2$  \\ \noalign{\bigskip}
$n_3(q_b)$ &
$32 \pi \frac{{\cal A}}{{\cal A}+1} \Gamma^2 \frac{\ln^3(q_b)}{q_b^6}$&
\begin{tabular}{r} 
$ \frac{4 \pi^2 c_{a}c_{b}}{q_b^5 \cosh\left(\frac{s\pi}{2}\right)}  \left\{\sqrt{\frac{{\cal A}}{{\cal A}+2}}\cos\left(s\ln \sqrt{\frac{{\cal A}+1}{2{\cal A}}}\right)\cosh\left[s\left(\frac{\pi}{2}-\theta_3\right)\right] \right.$ \\ 
$ \left. + \sin\left(s\ln \sqrt{\frac{{\cal A}+1}{2{\cal A}}}\right)\sinh\left[s\left(\frac{\pi}{2}-\theta_3\right)\right]\right\} $
\end{tabular}  \\ \noalign{\bigskip}
$n_4(q_b)$ &
$8 \pi \Gamma^2 \frac{\ln^3(q_b)}{q_b^6}$ &
\begin{tabular}{r} 
$\frac{ 8 \pi^2 |c_{a}|^2  }{s\; q_b^5 \cosh\left(\frac{s\pi}{2}\right)} 
\frac{{\cal A}^2}{\sqrt{{\cal A}({\cal A}+2)}}
  \left\{\sqrt{{\cal A}({\cal A}+2)}\sinh\left[s\left(\frac{\pi}{2}- \theta_4\right) \right] \right.$ \\
$\left. -\frac{s\; {\cal A} }{{\cal A}+1}\cosh\left[s\left(\frac{\pi}{2}- \theta_4\right) \right]\right\}$ 
\end{tabular} \\ \noalign{\bigskip}
$n_5(q_b)$ &
$\frac{32 \pi}{3} \frac{{\cal A} ({\cal A}-2)}{\left({\cal A}+1 \right)^2}  \Gamma^2 \frac{\ln^3(q_b)}{q_b^6}$ &
$-\frac{8\pi^2 \left|c_{a}\right|^2}{q_{b}^{5}} \frac{{\cal
A}^3({\cal A}+3)}{({\cal A}+1)^3 \sqrt{{\cal A}({\cal A}+2)}}$  \\
\noalign{\smallskip}\hline
\end{tabular}
\end{table}

In the following we discuss two interesting properties in the momentum distribution of $aab$ systems in both 2D and 3D. The geometric scaling of the Efimov states implies that observables may be described independently of the quantum state. The independence of the quantum states is not expected to be valid for 2D systems, since they do not present any geometric scaling. However, the leading order in the large momentum distribution was found to be independent of the state for three identical bosons \cite{bellottiPRA2013,wernerPRA2012} and for $aab$ systems, if the $aa$ subsystem is not interacting \cite{bellottiNJoP2014}.

\begin{figure}[!htb]%
\centering
\includegraphics[width=0.9\textwidth]{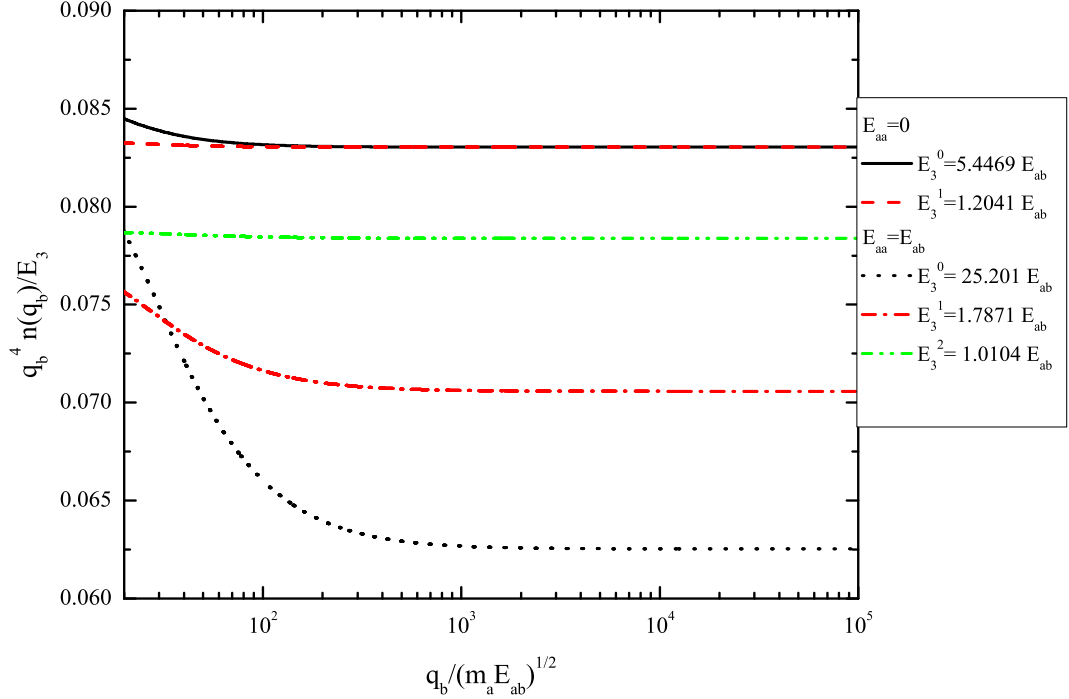}%
\caption[The leading order term of the one-body momentum density divided by $E_3^n$ for each
bound state labeled as $n$.]
{The leading order term of the one-body momentum density divided by $E_3^n$ for each
bound state labeled as $n$ in a system composed of two identical ($a=^{40}$K) particles
and a distinct one ($b=^{6}$Li) as a function of the momentum q for both $E_{aa}=E_{ab}$ and $E_{aa}=0$.}
\label{fig.07}%
\end{figure}

The effect of the two-body energy on the contact parameter is shown in Fig.~\ref{fig.07} for the $^{40}$K$^{40}$K$^{6}$Li system.  This system has three excited states when $E_{aa}=E_{ab}$ and only two when $E_{aa}=0$. Notice that the large momentum limit of the momentum density goes to a constant in all cases.  For $E_{aa}=0$ both momentum distributions are equal in units of the
three-body energy, i.e., $n^0(q_b)/E^0_3=n^1(q_b)/E^1_3$, where the superscript denotes the 
quantum state.  This case is rather special because the two identical particles
have zero energy and cannot provide a scale such that the three-body structure is determined 
by the identical two-body interactions in the identical subsystems.  In
other words the large-momentum limit of the one\=/body density for particle $a$ is determined
by the properties of the $ab$ subsystem. 

This picture changes when $E_{aa}=E_{ab}$, as seen in
Fig.~\ref{fig.07}. Now, in the large-momentum limit, the coefficients of the one\=/body densities change with the excitation energy.
The systematics is that the coefficients move towards the corresponding values for
$E_{aa}=0$ as function of excitation energy.  First the
differences of the ratios with the two\=/body energies is understandable, since the interaction of the two identical particles now must affect the three-body structure at small
distances, and hence at large momenta.  However, as the three-body
binding energy decreases, the size of the system increases and details
of the short-distance structure becomes less important. 

The independence of the state in the one-body momentum distribution of 2D three-body system can happen or not, depending on the energy of the two-body subsystem. However, both mass-imbalanced and identical bosons systems present the same functional form to the leading order (LO) and next-to-leading order(NLO) in the large momentum expansion of the one-body density. The same does not happens in 3D, since the NLO contribution due to non-oscillatory terms vanishes when ${\cal A}$=0.20, 1.00 and 1.57, i.e., the sum of the components $n_1(q_b),n_3(q_b),n_4(q_b),n_5(q_b)$ from table~\ref{tab:2} is null for these mass ratios \cite{yamashitaPRA2013}.   

We want to emphasize that both the results for 2D and 3D can be experimentally checked in the near future, since the calculations were made for alkali atoms. Besides the identical mass case, the other two mass ratios used in the 3D calculations can be obtained with mixtures of $^{133}$Cs and $^{87}$Rb (${\cal A}=1.565)$ or $^{6}$Li and $^{39}$K (${\cal A}=0.179)$.

\section{3D - 2D transition with PBC} \label{32pbc}

The physical and mathematical differences of three-body systems restricted to either 2D or 3D, presented in the previous sections, are the motivation that lead us to pursuit a method where the dimensionality enters as a parameter allowing to continuously interpolate between the well-known extremes of 2D and 3D. We restrict analyses to three-identical bosons, which presents the Efimov effect in 3D, but only two three-body bound states in 2D. Furthermore, the dimensionality plays an important role in the momentum distribution already in this simplest case, as can be seen in table~\ref{tab:2} for ${\cal A}=1$. 
 
Periodic boundary conditions (PBC) are assumed to be valid for the distance between the particles in the $z$ direction. The relative momentum is given by $\mathbf p_\perp=(p_x,p_y)$  in the flat 2D surface and by
\begin{align}
p_z=\frac{\sqrt{2} \pi n}{L}=\frac{n }{ R} \ , \;\;\; n=0,\pm1,\pm2, \hdots \; ,
\end{align}
in the transverse direction, with $L=\sqrt{2}\pi R$ being the size of the compact dimension corresponding
to a radius $R$, which is the parameter that dials between two and three-dimensions.
When $R \to 0$ it selects the 2D case and in the opposite limit,
i.e.,  $R\to \infty$, the 3D case is selected \cite{yamashitaJoPB2015}. 
The momentum $\mathbf{p}$ and its corresponding phase factor $d \mathbf{p}$ are, with PBC, defined as
\begin{align}
p^2= p_{\perp}^2+\frac{n^2}{R^2} \;\;\; \text{and} \;\;\; d\mathbf{p}=\frac{1}{R}d^2p_{\perp} \; .
\label{qdq}
\end{align}
We introduce the symbol $\sumint$, which indicates an integration over the continuum momentum in the plane ($p_{\perp}$) and a sum over the discrete perpendicular momentum ($p_z=\frac{n}{R}$). It reads
\begin{align}
\sumint d\mathbf{p} \equiv \sum_{n=-\infty}^{\infty} \int \frac{1}{R}d^2 p_{\perp} \; .
\label{sumintpbc}
\end{align}

Using definition~\eqref{qdq}, the three-body free Hamiltonian becomes
\begin{align}
H_0^p \left( \mathbf{q} , \mathbf{k} \right) &= \left( \mathbf{q}_{\perp} + \mathbf{q}_z \right)^{2} + \left( \mathbf{k}_{\perp} + \mathbf{k}_z \right)^{2} + \left( \mathbf{q}_{\perp} + \mathbf{q}_z \right) \cdot \left( \mathbf{k}_{\perp} + \mathbf{k}_z \right) \; , \nonumber\\
&= q_{\perp}^2+k_{\perp}^2+\mathbf{q}_{\perp} \cdot \mathbf{k}_{\perp} + \frac{n^2}{R^2}+\frac{m^2}{R^2} + \frac{n\;m}{R^2} \; 
\label{h0p}
\end{align}
and considering Eqs.~\eqref{qdq} to \eqref{h0p}, the integral equation for the bound state \eqref{spec} for a compact dimension with PBC  is found to be
\begin{multline}
f\left( \mathbf{q}_{\perp},n \right)   = - \frac{2}{R} \; \tau_p \left[ \frac{3}{4} \left(q_{\perp}^2+\frac{n^2}{R^2} \right)-E_3 \right]  
\sum_{m=-\infty}^{\infty} \int d^2k_{\perp}  \left( \frac{ f\left( \mathbf{k}_{\perp} , m\right)} {-E_{3}+H_0^p \left( \mathbf{q} , \mathbf{k} \right) }  - \frac{f\left( \mathbf{k}_{\perp} , m\right)}{\mu^2+H_0^p \left( \mathbf{q} , \mathbf{k} \right)  }\right)  \; ,
\label{bsecompac}
\end{multline}
with $\tau_p(E)$ given in Eq.~\eqref{taup} and $H_0^p \left( \mathbf{q} , \mathbf{k} \right)$ in Eq.~\eqref{h0p}. The subtraction is kept even after the discretization because the Thomas collapse is always present for any finite compact radius, no matter how small it is.
It is worthwhile to remind that, for $R \to \infty$, Eq.~\eqref{bsecompac} returns precisely the equation for the spectator function in 3D~\eqref{spec}. 

The two-body scattering amplitude  is 
\begin{align}
\tau_p(E)^{-1} = -\frac{2 \pi}{R} \ln \left[\frac{\sinh \left(\pi R \sqrt{|E|}  \right)}{\sinh \left(\pi R \sqrt{|E_2|}  \right)} \right] \; ,
\label{taup}
\end{align}
which recovers the matrix elements of 3D and 2D systems in the limits $R \to \infty$ and $R \to 0$, respectively. The first case is straightforward an reads
\begin{align}
\tau_{3D}^{-1}(E) = \lim_{R \to \infty} \tau_p^{-1}(E) 
= - 2\pi^2 \left(\sqrt{|E|} - \sqrt{|E_2|} \right)\; .
\label{tau3dpbc}
\end{align}
Going to the 2D limit, it is important to notice that a quasi-2D system is in practice a 3D system. Then, the units of  $\tau_{3D}^{-1}(E)$ and $\tau_p^{-1}(E)$ are exactly the same, but are different from $\tau_{2D}^{-1}(E)$. Taking into account the correct units, the 2D limit of Eq.~\eqref{taup} reads
\begin{align}
\tau_{2D}^{-1}(E) = \lim_{R \to 0}  R \; \tau_p^{-1}(E)  
= - 2\pi \ln \left(   \sqrt{\frac{|E|} {|E_2|}} \right) \; .
\label{tau2dpbc}
\end{align}
Expressions in Eqs.~\eqref{tau3dpbc} and \eqref{tau2dpbc} are  respectively identical to the expressions presented in table~\ref{tab:1} for 3D and 2D two-body T-matrix when $m_a=m_b=m_c$.

Introducing  dimensionless variables, $\epsilon_3=E_3/\mu^2$, $\epsilon_2=E_2/\mu^2$, $r=R\;\mu$, $y_\perp=q_\perp/\sqrt{\mu}$ and $x_\perp=k_\perp/\sqrt{\mu}$ and integrating over the angular dependence, since the focus is on states with zero angular momentum, the integral equation \eqref{bsecompac} is written as 
\begin{multline}
f\left( y_{\perp},n \right) =  \left\{ \pi \ln \left[\frac{\sinh \left(\pi r \sqrt{\frac{3}{4} \left(y_{\perp}^2+\frac{n^2}{r^2} \right)-\epsilon_3}  \right)}{\sinh \left(\pi r \sqrt{\epsilon_2}  \right)} \right] \right\}^{-1} \\* 
\times \sum_{m=-\infty}^{\infty} \int_0^{\infty}  dx_{\perp} \; x_{\perp} \; f\left( x_{\perp} , m\right) \left( \frac{1} {\sqrt{\left(-\epsilon_{3}+y_{\perp}^2+x_{\perp}^2+ \frac{n^2}{r^2}+\frac{m^2}{r^2} + \frac{n\;m}{r^2} \right)^2 - y_{\perp}^2\; x_{\perp}^2} } \right. \\* \left.
- \frac{1} {\sqrt{\left(1+y_{\perp}^2+x_{\perp}^2+ \frac{n^2}{r^2}+\frac{m^2}{r^2} + \frac{n\;m}{r^2} \right)^2 - x_{\perp}^2\; y_{\perp}^2} } \right)    \; .
\label{bsecompac1}
\end{multline}

\begin{figure}[!htb]
\centering
\includegraphics[width=0.9\textwidth]{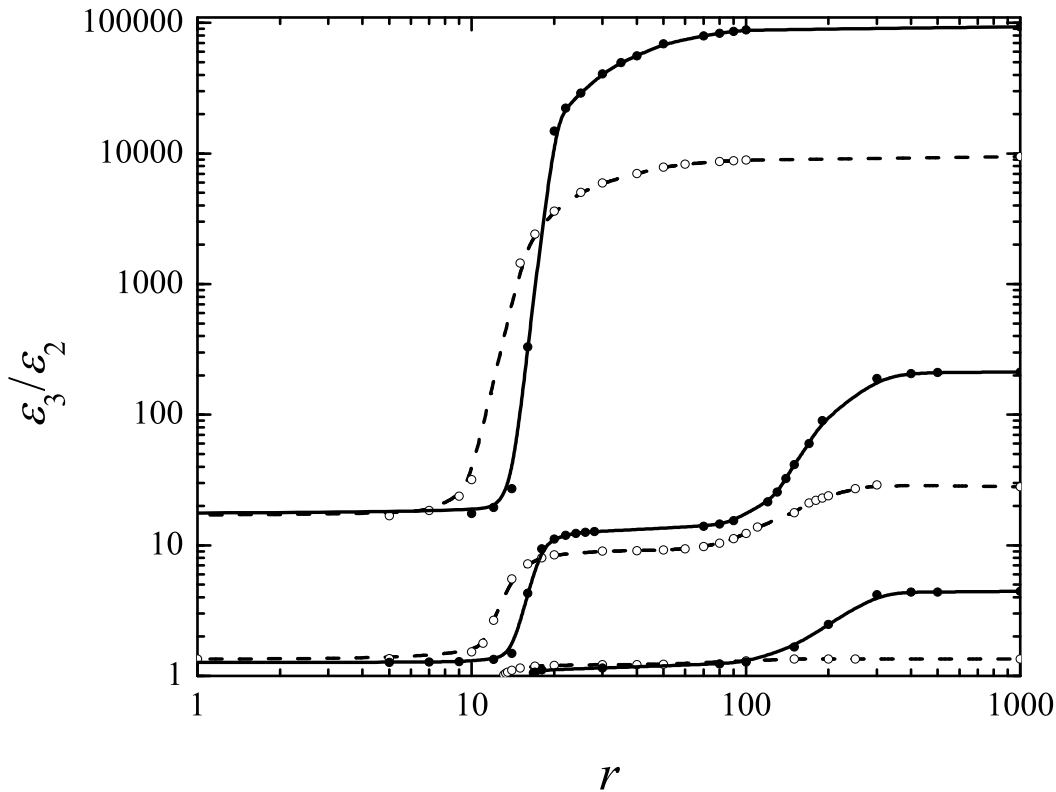}
\caption[Dimensional crossover of the three\=/body binding energy spectrum.]
{$\epsilon_3/\epsilon_2$ as a function of $r$, for $\epsilon_2=$ $10^{-7}$ 
(full circles) and $10^{-6}$ (empty circles). The solid and dashed lines are guides 
to the eye. As the 2D limit ($r\to 0$) is approached, higher 
excited states disappear and only the ground and first excited states remain.}
\label{compact}
\end{figure}

The dimensional crossover transition is explored through the numerical solution of Eq.~\eqref{bsecompac1}. In Fig.~\ref{compact} the ratios $\epsilon_3/\epsilon_2$ are showed as function of the compact dimension radius $r$, for the ground, first, and second
excited states. Notice that the last state goes into the continuum before the 2D limit is reached.

The computations were performed for two fixed two-body 
energies $\epsilon_{2}=10^{-6}$ (empty circles/dashed lines) 
and $10^{-7}$ (full circles/solid lines). Note that the Efimov 
ratio between two consecutive three-body states, $\sim 515$, 
is not completely reproduced for a finite $a$. 
The points at which the energies are calculated are 
showed explicitly, while the curves are guides to the eye and
for $r=1000$ the energies are obtained from the pure 3D equation.

An interesting dimensional crossover result is seen in Fig.~\ref{compact}, where only one sharp transition is present 
for the ground state while there are two for the first excited state. 
This behavior can be understood by
considering the size of the trimer given roughly by $\bar{r}\sim1/\sqrt{\epsilon_3}$. 
For $\epsilon_2=10^{-7}$, the ground state plateau for $\epsilon_3/\epsilon_2=93330$ is placed at $\bar{r}=10.35$  and first 
excited state plateau for  $\epsilon_3/\epsilon_2=211.79$ at $\bar{r}=217.29$.
These $\bar{r}$ values give 
approximately the region of the jumps signaling that the 3D limit, represented by the plateau, 
is reached once the trimer size matches the size of the squeezed dimension, $r$. The same 
analysis can be made for $\epsilon_2=10^{-6}$ with $\bar{r}=10.27$ and $\bar{r}=188.98$, 
respectively, for the ground and first excited state. Varying $r$ from large to small 
values, the 3D$\rightarrow$2D transition occurs for $r\sim10$, where it is possible to notice the 
disappearance of the higher excited states in order to reproduce the well known 
2D results with two trimer bound state energies proportional to $\epsilon_2$ with the ratios $\epsilon_3/\epsilon_2=16.52$ 
and $\epsilon_3/\epsilon_2=1.27$ \cite{bruchPRA1979}.

From the experimental point of view it may be difficult to keep the dimer energy 
constant. However, the transition observed in Fig.~\ref{compact} 
will not disappear due to a variation of $\epsilon_2$ with $r$. The increase of the 
dimer energy will merely move the beginning of the jumps towards smaller $r$. The 
optimal way to probe these jumps is to start from a two-body energy in  
the unitary limit ($a\to \infty$) where the 2D plateaus are fixed. Larger dimer energies will cause  
the 3D plateau to move to lower $\epsilon_3/\epsilon_2$ ratio and push the beginning of the transition 
to smaller $r$, thus making the transition region broader. Another interesting study about the dimensional crossover, where three identical bosons are confined by a harmonic potential along one direction is found in \cite{levinsenPRX2014}.

\section{Conclusion}

In this paper we summarized our main results involving the single particle momentum 
distributions in two and three dimensions. The summary tables, comparing the matrix elements 
and the functional form for the different terms of the momentum distribution makes the 
comparison between the 2D and 3D regimes much easier. In the last section we presented in 
a very schematical form a method to continuously interpolate between different dimensions. 
The same technique used here to go from 3D to 2D, may be used to go from 2D to 1D systems. 
All discussions in this text followed closely the papers \cite{bellottiNJoP2014,yamashitaPRA2013,yamashitaJoPB2015}.
An interesting direction for future investigation, which connects all the results discussed in this work, is to understand whether the contact parameter tell us how much of the wave function is in each dimension in the transition region presented in Fig.~\ref{compact}~\cite{valientePRA2012}.

\begin{acknowledgements}
We would like to thank Profs. Frederico, Zinner, Fedorov and Jensen for discussions. 
This work was partly supported by funds provided by the Brazilian agencies 
FAPESP (2013/04093-3), CNPq and CAPES (88881.030363/2013-01).
\end{acknowledgements}



\end{document}